\documentclass[preprint,12pt,sort&compress]{elsarticle}
\usepackage{amsmath}
\usepackage{graphicx}
\usepackage{subfigure}
\usepackage[latin1]{inputenc}

\newtheorem{teorema}{Theorem}
\graphicspath{{figuras/}{fig2/}}
\begin{document}

\begin{frontmatter}
\title{A nonlinear delayed model for the immune response in the presence of viral mutation}
\author[UFAL]{D. Messias}
\author[UFAL]{Iram Gleria}
\author[ARA]{S.S. Albuquerque}
\author[ARA,BU]{Askery Canabarro\fnref{t1}}
\fntext[t1]{Corresponding author, askery.canabarro@arapiraca.ufal.br}
\author[BU]{H. E. Stanley}

\address[UFAL]{Instituto de F\'\i{}sica, Universidade Federal de Alagoas, 57072-970 Macei\'o, Brazil}%
\address[ARA]{Grupo de F\'isica da Mat\'eria Condensada, N\'ucleo de Ciencias Exatas - NCEx, Campus Arapiraca, Universidade Federal de Alagoas, 53309-005 Arapiraca-AL, Brazil}
\address[BU]{Center for Polymer Studies and Department of Physics, Boston University, Boston, Massachusetts 02215, USA}

\begin{abstract}
We consider a delayed nonlinear model of the dynamics of the immune
system against a viral infection that contains a wild-type virus and a
mutant. We consider the finite response time of the immune system and
find sustained oscillatory behavior as well as chaotic behavior
triggered by the presence of delays. We present a numeric analysis and
some analytical results.
\end{abstract}


\end{frontmatter}


\section{Introduction}
We consider a nonlinear set of delay differential equations (DDEs) to
model the interaction of the immune system with an external pathogen,
e.g., a viral infection. Our model follows one presented in
Ref.~\cite{shu2014sustained} in which a time delay takes into account
the non-instantaneous immune response caused by a sequence of events
(e.g., activation of antigenic response or production of immune cells)
that occurs within a finite time period. In addition, the presence of
sustained aperiodic oscillations and chaotic trajectories observed in
real data \cite{askery,shu2012,ortiz2002} indicates that time delays are
needed to allow bifurcations that cause chaotic behavior even in models
that are one- and two-dimensional \cite{elder}. In ordinary differential
equations (ODEs) a minimum set of three coupled equations is required.

Because the fundamental underlying mechanisms are non-instantaneous,
several biological models have recently been modeled using delay
differential equations. Among these are a predator prey model with
delays \cite{smith2011introduction}, a model for the dynamics of the
hormonal control of the menstrual cycle \cite{clark2003bullet}, a model
for human respiration \cite{batzel2001AMC}, a model for dioxide carbon
levels in the blood \cite{culshaw2000delay,wang2006viral}, and a number
of models for viral dynamics
\cite{askery,elder,iram,shu2014sustained,nelson2002mathematical,tam,shi,culshaw,buric,song,liu,iram2,culshaw2,bocharov}.

In previous research \cite{askery} we analyzed the cellular immune
response and found that stationary solutions bifurcate to an unstable
fixed point when delays are longer than a critical immune response time
$\tau_c$. We found that increasing the time delay causes the system to
suffer a series of bifurcations that can evolve into a chaotic regime.
We used two coupled delayed equations to model the interaction of the
immune system with a target population \cite{elder}. We used some
analytical tools to analyze delayed systems \cite{iram}, and we
published new results for the model originally presented in
Ref.~\cite{askery}. Here we consider a three-dimensional version of a
model that previously appeared in the literature
\cite{shu2014sustained,komarova2003boosting} for the dynamics of the
population of virus $y(t)$ and of immune cells $z(t)$, and also a mutant
population of virus $y_m(t)$.

Delay differential equations require both the initial conditions and the
history of the dynamic variable values of $t<\tau$. Because we are using
models with discrete delays, $\tau$ is constant. This is in contrast to
a system with distributed delays in which $\int_{t-r}^{t} k(t-s)x(s) ds
= \int_0 ^rk(z)x(t-z)dz$, where $0\leq r \leq \infty$ is the distributed
delay and the kernel $k$ is normalized, and thus $\int_{-\infty}^\infty
k(y)dy = 1$. For an identically null $k(u),\forall u>u_{\rm max}$ the
delay can be represented by integrals of type $\int_{- \infty}^{t}
M_{1}(s)k(t-s) ds = \int_0 ^\infty M_{1}(t-u)k(u)du$. These are
``bounded delays'' because they represent the values of $M_{1}$ at a
past time $(t-u_{\rm max},t)$. A discrete delay is a particular kind of
bounded delay. More complicated forms are also possible, e.g., delays of
type $x(t-r[x(t)])$ distributed over space.

Introducing delays allows us to model richer behavior, e.g., the
well-known logistic equation governing the dynamics of a population
density $N(t)$: $\dot{N}(t)= N(t) \left(1 - \frac{N(t)}{K} \right)$,
with $r$ the growth rate and $K$ the carrying capacity. Note that for
every initial condition $N(0)$ the system ultimately reaches the stable
equilibrium $N(t)\rightarrow K$. A delayed version of this model can be
used for a species population that gathers and stores food, i.e., when
resources vanish, the species population starves within finite time
$\tau$. Reference~\cite{hutchinson1948circular} assumes this and
analyzes the delayed system $\dot{N}(t)= N(t)r \left(1 -
\frac{N(t-\tau)}{K} \right)$.  This delayed version of the logistic
equation can model chaotic behavior that instantaneous one dimensional
models cannot because ODE systems need at least a three-dimensional
state space to model chaos, as demonstrated in Lorenz's seminal work
\cite{lorenz1963}. Here the number of initial conditions is equal to the
number of degrees of freedom. In delayed systems the number of degrees
of freedom is infinity and chaos occurs in even one dimensional systems,
as in the case for one-dimensional non-invertible maps.

We present the model in the next section. In section \ref{anali} we
present some analytical and numeric results, and in section \ref{conc}
we present our conclusions.

\section{Model}
\noindent
Our model is based on research described in
Refs.~\cite{shu2014sustained,komarova2003boosting} that uses a
two-dimensional model for the dynamics of the population of virus $y(t)$
and of immune cells $z(t)$. We use time-lagged response for the immune
system, following previous research demonstrating its importance in the
appearance of the Hopf bifurcations \cite{shu2012}, chaotic trajectories
\cite{askery}, and sustained oscillatory behavior rarely seen in the
instantaneous version of the model \cite{ortiz2002}. Here we extend the
model to a spreading population of mutant virus $y_m(t)$,
\begin{eqnarray}\label{modelo}
\dot{y}&=&r (1-\alpha)y(t) \left(1-\frac{y(t)}{K} \right) - ay(t) -
py(t)z(t)\\ \nonumber 
\dot{y_m}&=&\alpha_m r_m y_m(t) \left(1-\frac{y_m(t)}{K_m}
\right)-a_my_m(t) - p_my_m(t)z(t)\\ \nonumber 
\dot{z}&=&\frac{cy(t-\tau_1)z(t-\tau_1)}{1+dy(t-\tau_1)} +
\frac{c_my_m(t-\tau_2)z(t-\tau_2)}{1+d_m y_m(t-\tau_2)} - qy(t)z(t) -
q_my_m(t)z(t) - bz(t), 
\end{eqnarray}

where $r(1-\alpha)$ is the growth rate of the viral population for
$y\approx 0$. This rate decreases and reaches zero when the virus
population equals $K$. The virus population decays with $a$. We then
have a net rate of $r(1-\alpha)-a$ and a carrying capacity of
$\frac{K(r(1-\alpha)-a)}{r(1-\alpha)}$. The viruses are eliminated by
cells of the immune system according a rate $p$. The term $y_m$
represents the concentration of the mutant viruses. Its net growth rate
and carrying capacity are, respectively, $r_m\alpha_m-a_m$ and
$\frac{K_m(r\alpha_m-a)}{r\alpha_m}$. They are eliminated at a rate
$p_m$.  The immune cell concentration $z$ grows proportionally to the
virus population according to a saturation term. The $\tau_2$ value is
the delay in the immune response to the viral infection. The delay
$\tau_1$ refers to the processes used by the organism to prepare the
cells to fight the virus. Immune cells are attacked and destroyed by the
original viruses and their mutant version with rates $q$ and $q_m$,
respectively.  The terms $1/(1+dy(t-\tau_1))$ and $1/[1+d_m
  y_m(t-\tau_2)]$ shows that the immune response is proportional to the
product of the virus (either $y$ or $y_m$) and the population of immune
cells $z$, but saturates when the virus population is higher.  Numerical
estimations of the parameters are provided in
Ref.~\cite{komarova2003boosting}.

\begin{center}
\begin{tabbing}
\hspace{3cm}\=\hspace{5cm}\=\kill $r = 6 $ day$^{-1}$,\> $K=3$virus
mm$^{-3}$, \> $p = 1$ mm$^3$ cells $^{-1}$ day$^{-1}$, \\ $a=3$
day$^{-1}$ , \> $c=4$ mm$^3$ virus $^{-1}$ day$^{-1}$, \> $d=0.5$ mm
$^{-3}$ virus $^{-1}$, \\ $b=1$ day$^{-1}$, \> $q = 1$ mm$^{-3}$ virus
$^{-1}$ day$^{-1}$.  \>
\end{tabbing} 
\end{center}

Identical numeric values are assumed for $K_m$, $r_m$, $a_m$, $c_m$,
$d_m$, and $q_m$. The $p_m = 0.9 < p$ value is an exception because here
it is more difficult for the immune system to eliminate cells infected
by the mutant virus. We also assume $\alpha=1$ and $\alpha_m=0.05$,
which indicates that the mutation is a residual portion of the
replication mechanisms.

\section{Results} \label{anali}

\noindent
Reference \cite{shu2014sustained} presents several analytical results
for the two-dimensional version presented in (\ref{modelo}), which
does not take into account the mutant population $y_m$. Because our
model is three-dimensional it is cumbersome to analyze, and we focus on
numeric results. Similar to the procedure used in the logistic map, we
focus on the emergence of bifurcations and chaos as time-delay values
increase. The system in (\ref{modelo}) has a total of 11 equilibrium
points. Six are facial points (with at least one null component). Those
with simple algebraic expressions are
\begin{small}
\begin{eqnarray}
y&=&0,y_m=0,z=0; \nonumber \\
y&=&0,y_m=\frac{K_m(r_m-a_m)}{r_m},z=0; \nonumber \\
y&=&\frac{K(\alpha r+a-r)}{(r(-1+\alpha)}, y_m = 0, z = 0. \nonumber \\
\end{eqnarray} \label{fix}
\end{small}

\noindent
The others present cumbersome algebraic expressions, which we omit
here for sake of simplicity.
 
Note that the stability of the fixed points of a $n$-dimensional system
with $k$ delays can be analyzed using the usual Jacobian evaluated at
the equilibrium point \cite{iram}. Each $\dot{x_i},i=1,\cdots,n$ be
written
$${\dot x_i}=\sum_{j=1}^{k}F_j^i\left(x_1(t-\tau_j),x_2(t-\tau_j),\cdots\right).$$ 
Performing a series expansion around the equilibrium point
$x^*=(x_1^*,\cdots,x_n^*)$, we obtain for each $x_i,i=1,\cdots,n$
$${\dot x_i} \approx \sum_{j=1}^{k}\left( F_j^1(x_1,\cdots)|_{x^*}+
\frac{\partial F_j^i}{\partial x_1}|_{x^*}(x_1(t-\tau_j)-x_1^*) +\frac{\partial F_j^i}{\partial x_2}|_{x^*}(x_2(t-\tau_j)-x_2^*)+\cdots \right).$$
We then have a linear system of variables $\tilde{x}_i\equiv x_i-x_i^*$
with $k$ Jacobian matrices that take the form
\begin{equation}
J_j=\left[
\begin{array}{ccc}
\frac{\partial F_j^1}{\partial x_1} & \frac{\partial F_j^1}{\partial
  x_2} & \cdots   \\ 
\vdots & \vdots & \vdots  \\ 
\frac{\partial F_j^n}{\partial x_1} & \frac{\partial F_j^n}{\partial
  x_2} & \cdots \\ 
\end{array}
\right];
\end{equation}
evaluated at the fixed points. The stability of a particular fixed point
is determined by the eigenvalues of its corresponding
Jacobian. Bifurcations occur whenever one eigenvalue crosses the
imaginary axis as one or more parameters, including the delays,
change. Typical bifurcations involve a turning point when the eigenvalue
is initially null, and a Hopf bifurcation when a pair of complex
eigenvalues crosses the imaginary axis \cite{iram}.

The general expression for the Jacobian is
\begin{tiny}
\begin{equation} \label{btil}
J=
\begin{bmatrix}
r(1-\alpha)(1-\frac{2y^*}{K})-a -pz^*& 0 & -py^*\\
0 & \alpha_mr_m(1-\frac{2y_m}{K_m})-a_m-p_mz^* & -p_my^*_m\\
\left(\frac{cz^*}{(1+dy^*)^2} -qz^*\right)e^{-\lambda \tau_1} &
\left(\frac{c_mz^*}{(1+d_my_m)^2}-q_mz^*\right)e^{-\lambda \tau_2} &
\frac{cy^*}{(1+dy^*)}+ \frac{c_my^*_m}{(d_my^*_m+1)}-qy^*-q_my^*_m-b\\ 
\end{bmatrix}
\end{equation}
\end{tiny} 
The Jacobian for the origin is thus
\begin{equation}
\tilde{J}=\left[ 
\begin{array}{ccc}
r(1-\alpha)-a & 0 & 0   \\ 
0 & r_m-a_m & 0 \\ 
0 & 0 & -b 
\end{array}
\right]. , \label{Jt}
\end{equation}

\begin{figure}[!htb]
\begin{center}
\subfigure[]{\includegraphics[width=0.45\textwidth]{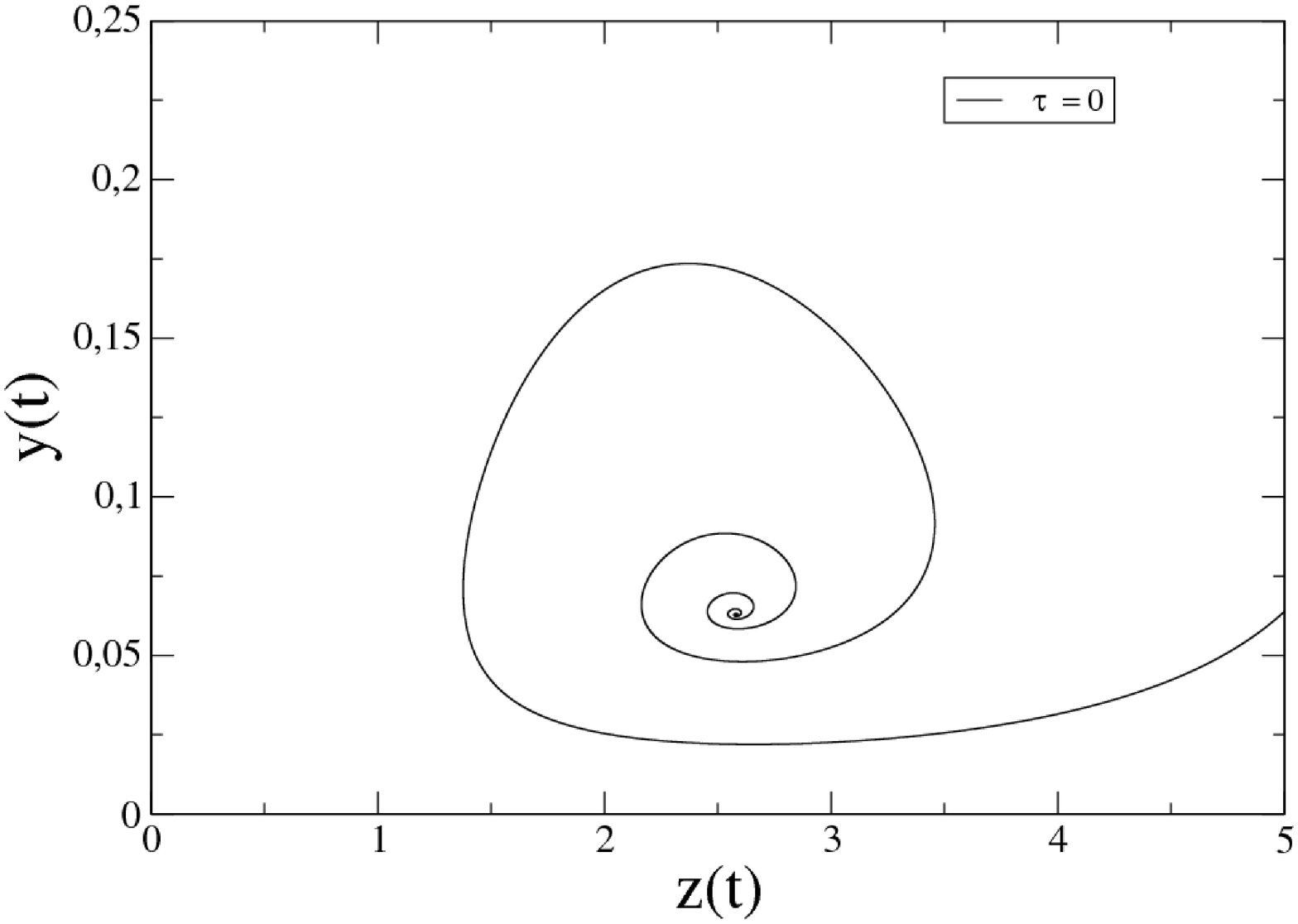}
}
\quad
\subfigure[]{\includegraphics[width=0.45\textwidth]{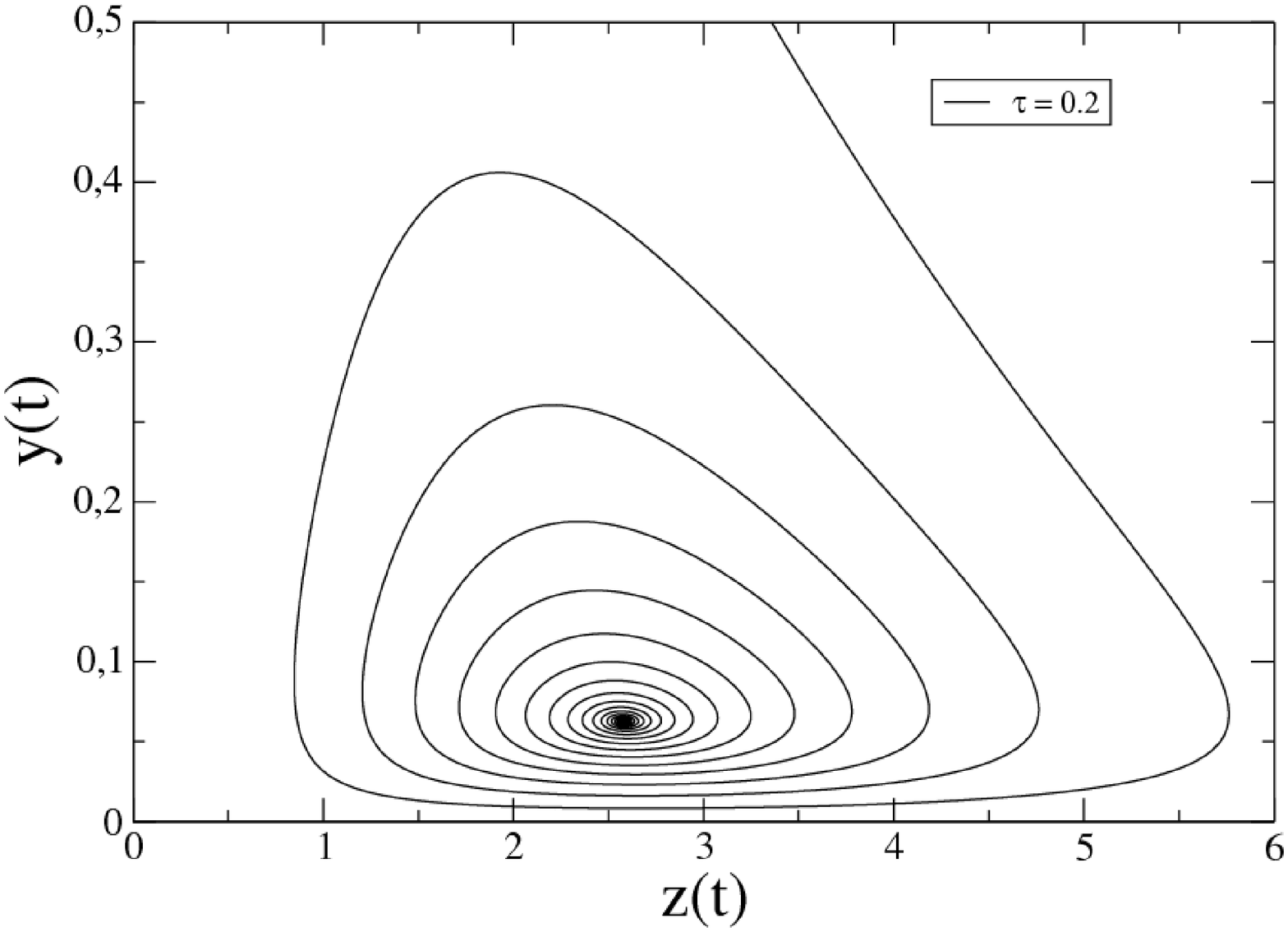}
}
\subfigure[]{\includegraphics[width=0.45\textwidth]{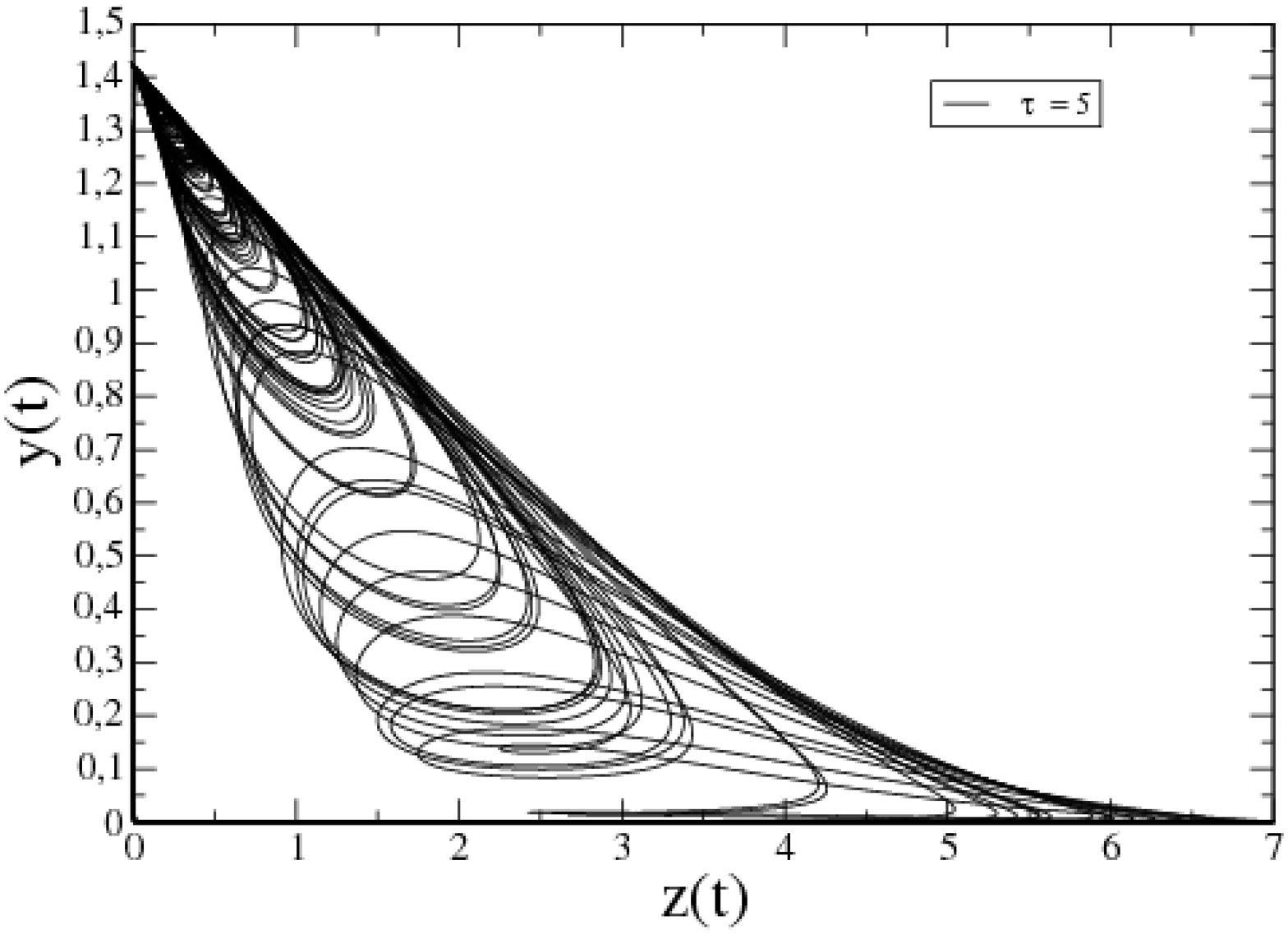}
}
\quad
\subfigure[]{\includegraphics[width=0.45\textwidth]{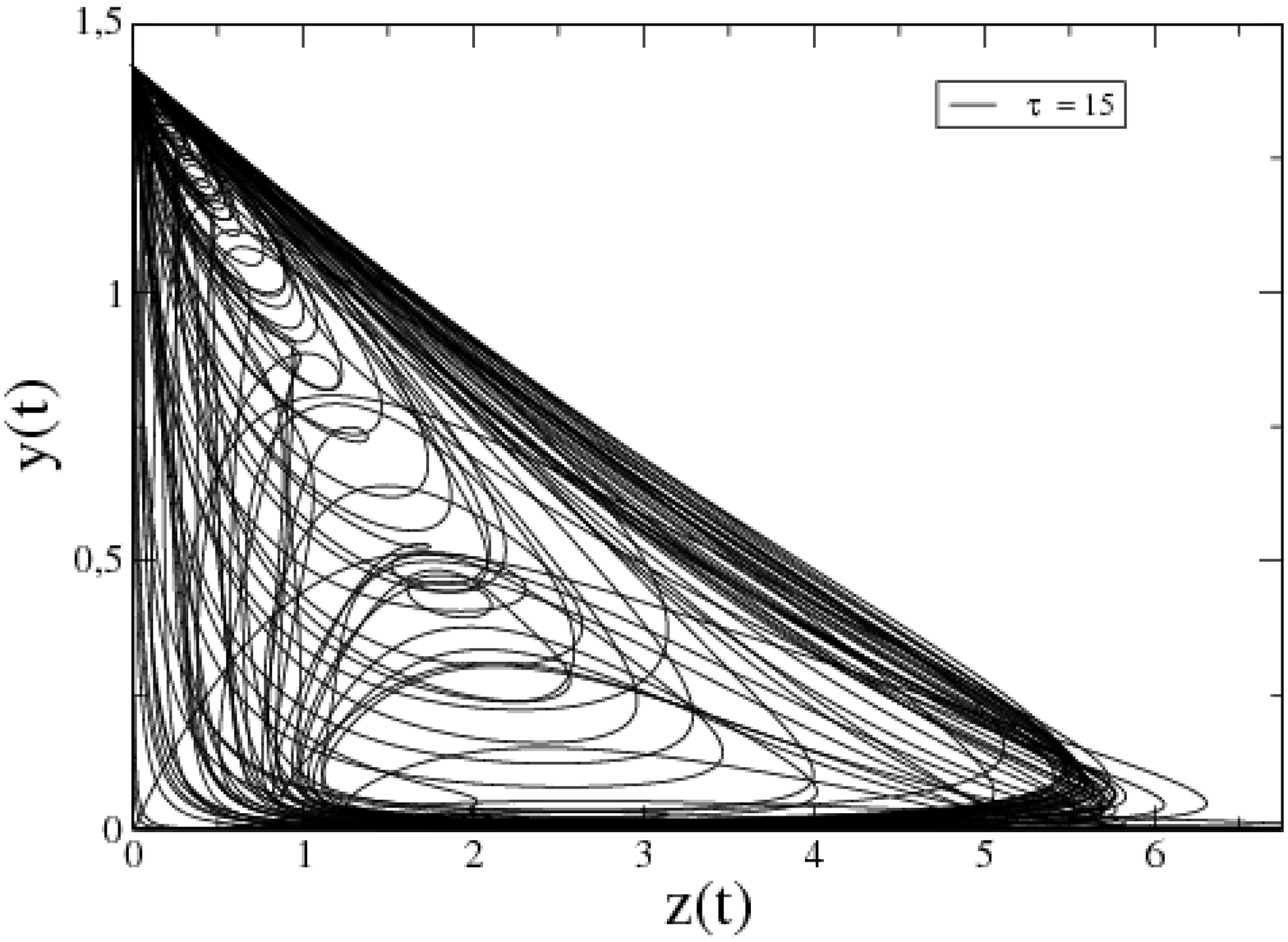}
}
\caption{Phase portraits for the case $\tau_1=\tau_2= \tau$. In (a) we
  have $\tau=0$. In (b) $\tau=0.2$, $\tau=5$ in (c) and ($\tau = 15$) in
  (d).}
\label{fig1}
\end{center}
\end{figure} 

which holds for all values of $\tau_1,\tau_2$. The eigenvalues are
$-\alpha r-a+r$, $r_m-a_m$, and $-b$. Stability (with only negative
eigenvalues) can be achieved for smaller $r$ and $r_m$ (the viral growth
rate), and larger $a,a_m$ (the natural population decay of the
virus). Here ultimately the system loses all of its viruses and has no
immune cells irrespective of the delay. For the set of parameters chosen
here, however, the origin is unstable $\forall \tau_1,\tau_2$. Note that
for the three fixed points in (\ref{fix}) the stability is unchanged
when there are non-null delays. This can be seen from (\ref{btil}) by
substituting $z=0$. Note that $r-\alpha$, $r-a$, and $-r_m+a_m$ are
common eigenvalues, a condition that renders the origin unstable for all
three.

\begin{figure}[!h]
\begin{center}
\includegraphics[width=0.45\textwidth]{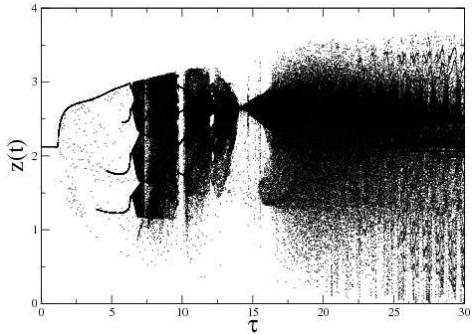}
\end{center}
\caption{The same model considered in \cite{shu2014sustained}, with the
  same set of parameters used in our model (\ref{modelo}). Some windows
  of regular behavior are observed.}
\label{fig2}
\end{figure} 

For null delays and the chosen set of parameters, only two of the 11
equilibria are stable. Because $y,y_m$ and $z$ are densities and
therefore positive quantities, one stable equilibrium is physically
irrelevant: $y = 5.568989996$, $y_m = 5.046486447$, and $z =
-7.88108099$. The other stable equilibrium is the spiral focus (SF): $y
= 0.06265629108$, $y_m = 0.3385711289$, and $z = 2.580953047$. For the
parameters used, the remaining equilibria are all unstable and comprise
six facial equilibria and two physically-irrelevant equilibria that have
at least $y<0$ or $y_m<0$ or $z<0$. Here we focus on how increasing the
value of the time delay alters the stability of the stable SF solution.

\begin{figure}[!htb]
\begin{center}
\includegraphics[width=0.45\textwidth]{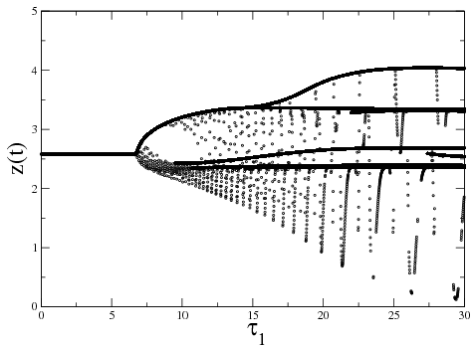}
\end{center}
\caption{Bifurcation diagram in function of $\tau_1$ for 
  $\tau_2=0$. Chaotic behavior is not observed.}
\label{fig3}
\end{figure}

\begin{figure}[!htb]
\begin{center}
\subfigure[]{\includegraphics[width=0.45\textwidth]{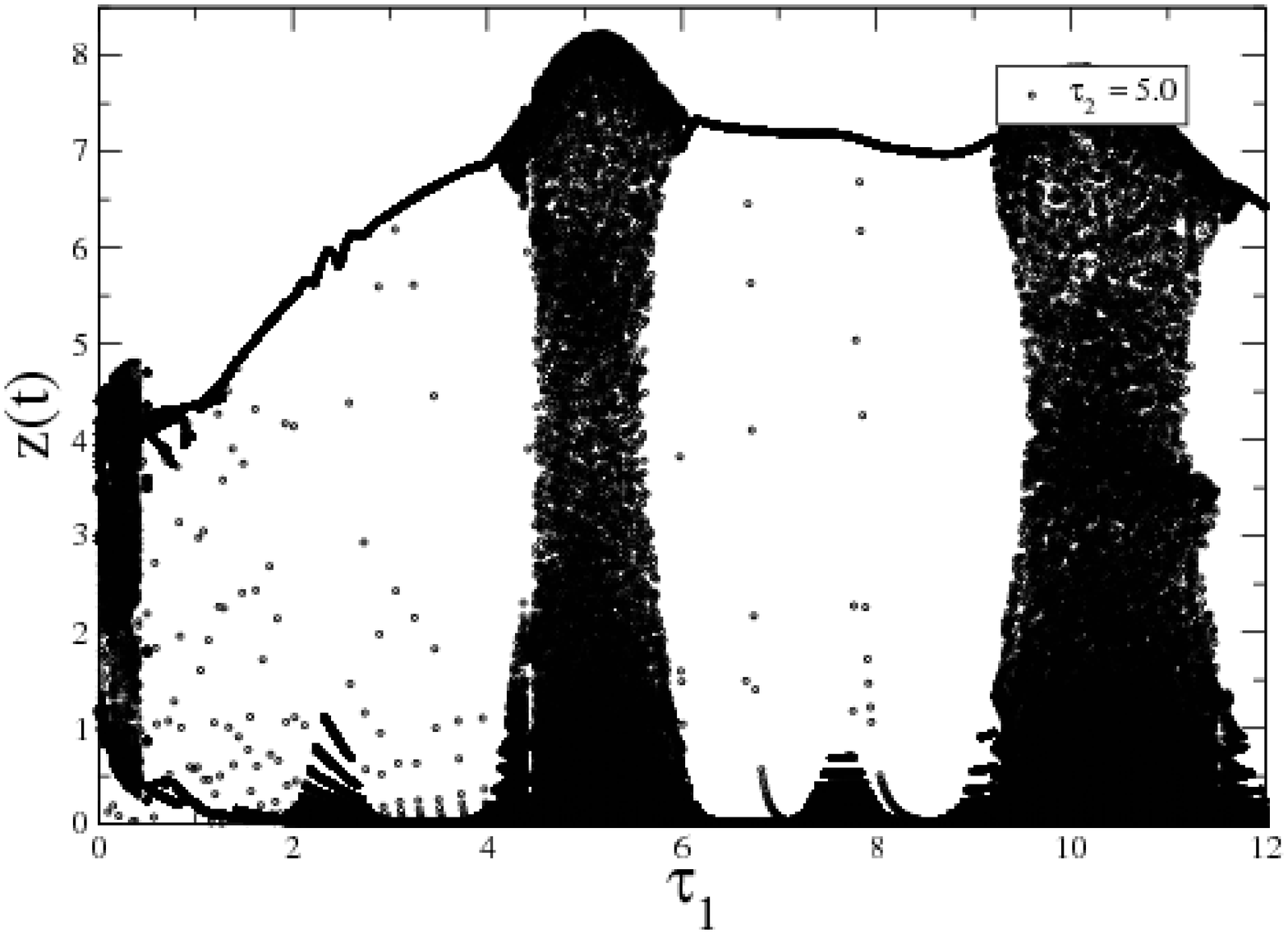}
\label{fig3a}}
\subfigure[][]{\includegraphics[width=0.45\textwidth]{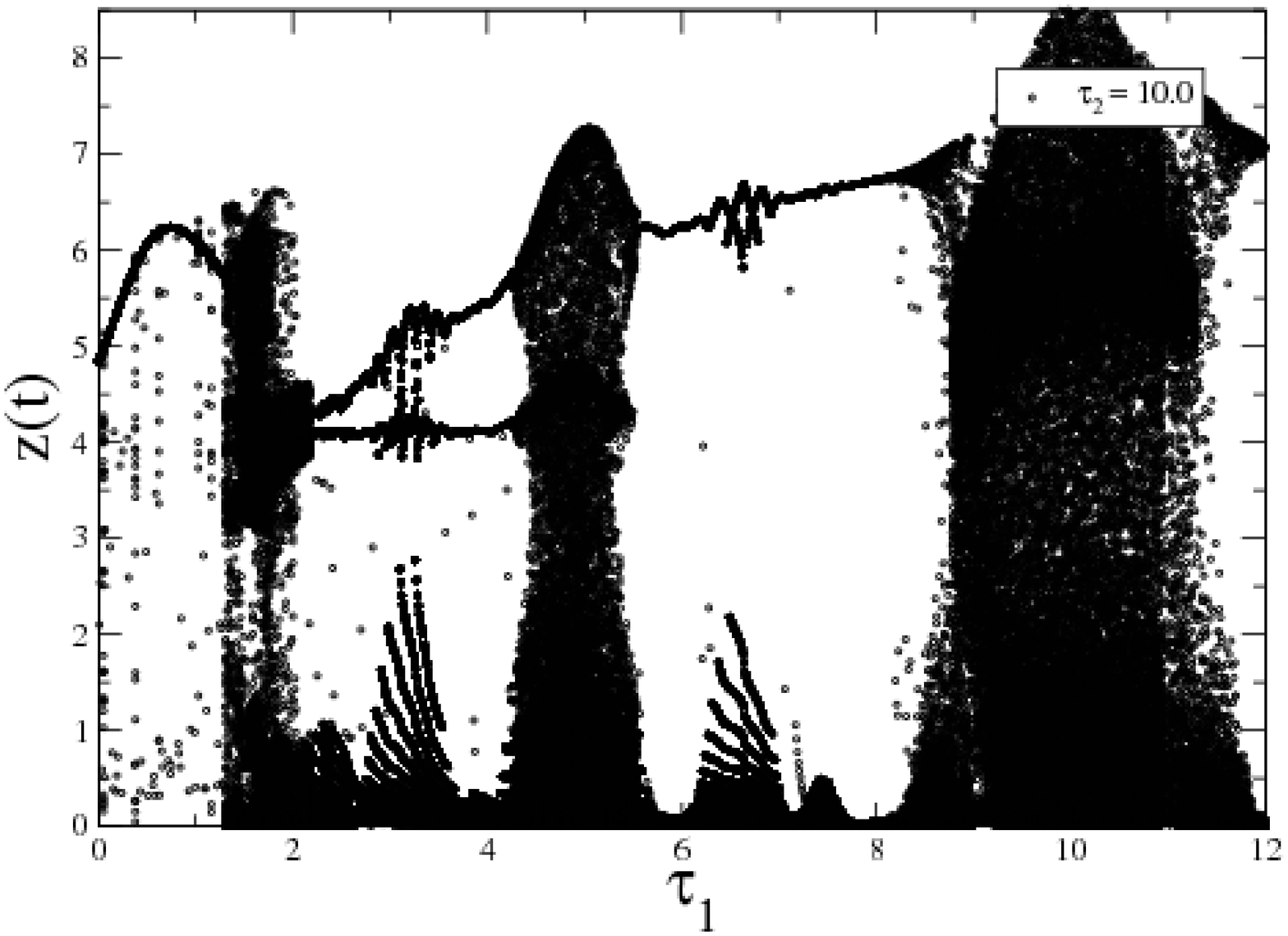}
\label{fig3b}}
\subfigure[]{\includegraphics[width=0.45\textwidth]{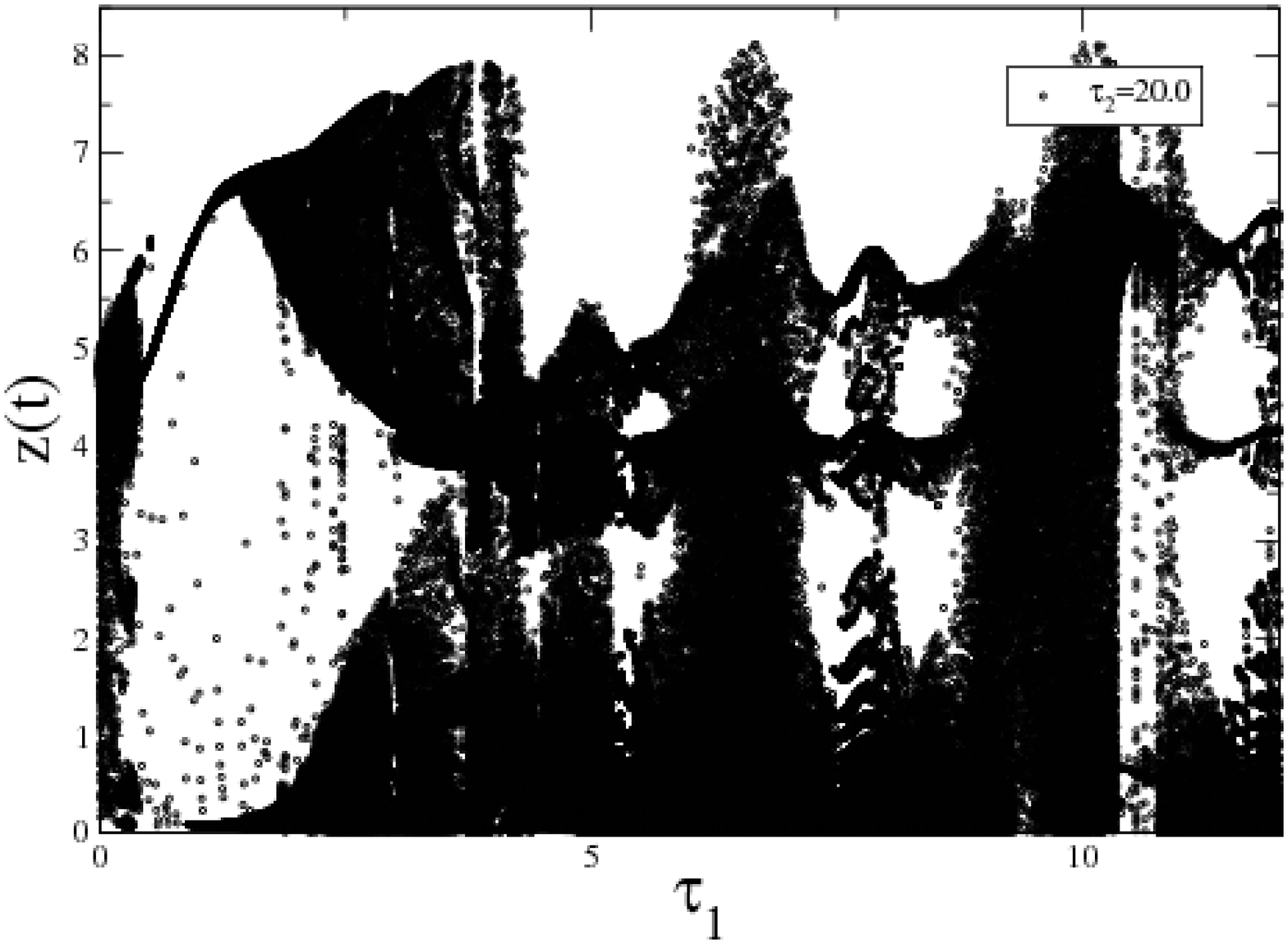}
\label{fig3d}}
\caption{ The maximum $z(t)$ as function of $\tau_1 $ for fixed
  $\tau_2$. The presence of $\tau_2$ allows the emergence of chaotic
  behavior.  In \ref{fig3a}, $\tau_2=5$. In \ref{fig3b}, $\tau_2=10$.
In \ref{fig3d}, $\tau_2=20$}
\end{center}
\label{fig4}
\end{figure}

A theorem presented in Ref.~\cite{vinte} describes the conditions for
switches in stability when there are delays and finds a critical
$\tau^*>0$ above which the equilibrium point is always unstable. The
theorem states:

\begin{teorema}

Let a characteristic equation of a given fixed point be written
$R(\lambda)+S(\lambda)\exp(-\lambda\tau)=0$. $R(\lambda)$ and
$S(\lambda)$ are analytical in the right half plane and
$\Re{\lambda}>-\delta,\delta>0$. When the following properties hold:

\begin{itemize}

\item[{(i)}] $R(\lambda)$ and $S(\lambda)$ have no common zero;

\item[{(ii)}] $\overline{R(-Is)}=R(Is)$,$\overline{S(-Is)}=R(Is)$, where
  the bar indicates the conjugate and $I=\sqrt(-1)$;

\item[{(iii)}] $R(0)+S(0)=0$;

\item[{(iv)}] The half right plane possesses at most a finite number of
  roots of $R(\lambda)+S(\lambda)\exp(-\lambda\tau)=0$ when $\tau=0$;
  and

\item[{(v)}] $F(y)=\left|R(Iy)\right|^2-\left|S(Iy)\right|^2$ when real
  $y$ has at most a finite number of zeros.

\end{itemize}

Then the following statements are true:

\begin{itemize}

\item[{(a)}] If $F(y)=0$ has no positive real roots, and if the
  associated fixed point is stable (unstable) for null delays, it will
  remain stable (unstable) for all delays.

\item[{(b)}] If $F(y)=0$ has at least one positive root and all roots
  are simple, stability switches can occur with increasing $\tau$. There
  exists a $\tau^*>0$ above which the fixed point is unstable for all
  $\tau>\tau^*$. As $\tau$ varies from zero to $\tau^*$ at most a finite
  number of stability switches may occur.

\end{itemize}

\end{teorema}

Reference \cite{elder} used this theorem to analyze their model. Here we
consider equal delays $\tau_1=\tau_2=\tau$. The Jacobian of the spiral
focus stable equilibrium ($y = 0.06265629108$, $y_m = 0.3385711289$, $z
= 2.580953047$ is
\begin{eqnarray}
&J_{SF} =
  (-0.4825880248-1.960851071\lambda)
\exp(-\lambda\tau)-0.7961892098\lambda^{2}-\lambda^{3}
  \nonumber \\  
&-0.08061172163\lambda+8.061172243\dot 10^{-11}. 
\end{eqnarray}
This equation is clearly of type
$R(\lambda)+S(\lambda)\exp(-\lambda\tau)$. Thus
$F(Iy)=\left|R(Iy)\right|^2-\left|S(Iy)\right|^2$ yields
\begin{eqnarray}
F_{SF} = 0.4726938145y^4+y^6-3.838438673y^2-0.2328912017.
\end{eqnarray}
The roots of this equation are $\pm 1.330454624, \pm 0.2455259061I, \pm
1.477335558I$, and thus the condition $b$ of item $v$ of the theorem
holds. Figure~\ref{fig1} shows the expected stability switches. We plot
$z(t)$ versus $y(t)$ for $\tau_1=\tau_2=0$ (the stable case),
$\tau_1=\tau_2=0.2$, $\tau_1=\tau_2=5$, and $\tau_1=\tau_2=15$. There is
still stability for delay $\tau = \tau_1=\tau_2=0.2$, but this is lost
in $\tau = 5$ and $\tau = 15$. These results demonstrate how the
introduction of delays can change the stability of a stable solution and
promote a richer dynamics for the system.

For the sake of comparison, we use the two-dimensional model proposed in
Ref.~\cite{shu2014sustained} (which has no mutant virus) and plug
$y_m=0$ and $K_m, r_m a_m,c_m,d_m,q_m,\tau_2=0$ into
(\ref{modelo}). Figure~\ref{fig2} shows the maxima values of $z(t)$
versus $\tau_1$. Note that there is a series of bifurcations that
switches between sustained oscillations and chaotic behavior, with
windows of periodic behavior (e.g., around $\tau_1=14$).

When we use the term $\frac{c_my_m(t)z(t)}{1+d_m y_m(t)}$ with
$\tau_2=0$ to introduce the mutant component into our model, it changes
the dynamics of (\ref{modelo}). Figure~\ref{fig3} shows that periodic
orbits are present but not chaotic behavior. Although merely inserting a
new equation into the system does not enrich the dynamics, the situation
changes completely when $\tau_2\neq 0$. Figure~\ref{fig4} shows that
this time delay causes more complex patterns to emerge, including
regions of chaotic behavior.

\section{Conclusion} \label{conc}

\noindent
We have considered a nonlinear set of delay differential equations to
model the interaction between an immune system and an external pathogen,
e.g., a viral infection. We extend the previous model considered in
\cite{shu2014sustained} by introducing a new variable that takes into
account mutant viruses. We find a series of bifurcations that lead to
chaotic behavior, an outcome that agrees with the results observed in
real data \cite{askery,shu2012,ortiz2002} and that corroborates previous
work indicating the need for the time delays that generate richer
behavior \cite{shu2014sustained}.

\section*{Acknowledgments}

\noindent
AC thanks the Alagoas State Research Agency FAPEAL for support through
major projects (PPP - 20110902-011-0025-0069 / 60030-733/2011), also
CNPq for PDE (207360/2014-6) and Universal (423713/2016-7) grants. DM
acknowledges a scholarship by the Brazilian funding agency CAPES. The
Boston University work was supported by DTRA Grant HDTRA1-14-1-0017, by
DOE Contract DE-AC07-05Id14517, and by NSF Grants CMMI 1125290, PHY
1505000, and CHE-1213217.

\end{document}